\magnification=1200
\hoffset=.0cm
\voffset=.0cm
\baselineskip=.55cm plus .55mm minus .55mm

%%%%%%%%%%%%%%%%%%%%%%%%%%%%%%%%%%%%%%%%%%%%%%%%%%%%%%%%%%%%%%%%%%%%%%%%%%%%%%%
%
%       Font loading
%
%	The following makes the AmS fonts available
%	(needs the files amssym.def and amssym.tex)
%
\input amssym.def
\input amssym.tex
%
% 
%%%%%%%%%%%%%%%%%%%%%%%%%%%%%%%%%%%%%%%%%%%%%%%%%%%%%%%%%%%%%%%%%%%%%%%%%%%%%%%
%
%
%  Qui di seguito si effettua la definizione della gerarchia matematica
%  per i fonti greci grassetti e per i fonti sans-serif in modo che 
%  venga compiuta la riduzione di corpo a esponente e a pedice quando
%  siano utilizzati in formule matematiche. 
%
%  Modifica effettuata da GS in data  28 novembre 1996
%
%  (Si utilizzano le famiglie numero 13 e numero 14 rispettivamente
%   sperando che non si vada in conflitto con altre famiglie gia` in uso)
%
%                  NON CAMBIARE SENZA GIUSTIFICATO MOTIVO
%                  E/O  SENZA IL PARERE DELL'AUTORE DELLA 
%                  MODIFICA  PRESENTE.
%
%

%%%%%%%%%%%%%%%%%%%%%%%%%%%%%%%%%%%%%%%%%%%%%

\font\grassettogreco=cmmib10
\font\scriptgrassettogreco=cmmib7
\font\scriptscriptgrassettogreco=cmmib10 at 5 truept
\textfont13=\grassettogreco
\scriptfont13=\scriptgrassettogreco
\scriptscriptfont13=\scriptscriptgrassettogreco

% Definition of sansserif fonts

\font\sansserif=cmss10
\font\scriptsansserif=cmss10 at 7 truept
\font\scriptscriptsansserif=cmss10 at 5 truept
\textfont14=\sansserif
\scriptfont14=\scriptsansserif
\scriptscriptfont14=\scriptscriptsansserif

% Definition of script fonts

\font\capital=rsfs10
\font\scriptcapital=rsfs10 at 7 truept
\font\scriptscriptcapital=rsfs10 at 5 truept
\textfont15=\capital
\scriptfont15=\scriptcapital
\scriptscriptfont15=\scriptscriptcapital
\def\scri{\fam=15}

% Definition of Euler  fonts

\font\euler=eusm10
\font\scripteuler=eusm7
\font\scriptscripteuler=eusm5 
\textfont12=\euler
\scriptfont12=\scripteuler
\scriptscriptfont12=\scriptscripteuler

%

%

%
%   FINE DEFINIZIONE DELLE FAMIGLIE MATEMATICHE PER I FONTI
%   cmmib10 (greci grassetti)  e  cmss10  (sans-serif)
%
%
%---- qui di seguito c'e` un campione di come vanno utilizzati ----     
%---- i diversi caratteri ESCLUSIVAMENTE IN MODO MATEMATICO.  ----
%---- Nelle definizioni l'unica cosa che puo` essere cambiata  ----
%---- ad arbitrio dell'utente e` il nome delle macro           ----
%---- (e.g. \Gammabf). Tutto il resto va lasciato come sta.   ----
%
%---- Se si vogliono definire anche i caratteri sansserif va  ----
%---- sostituito ovunque \bgr  con \ssm.                      ----
%

%
%
%        End of font calling
%
%%%%%%%%%%%%%%%%%%%%%%%%%%%%%%%%%%%%%%%%%%%%%%%%%%%%%%%%%%%%%%%%%%%%%%%%%%%%%%
%
%
%                           This is for referencing
%
\def\ref#1{\lbrack#1\rbrack}
%
%
%%%%%%%%%%%%%%%%%%%%%%%%%%%%%%%%%%%%%%%%%%%%%%%%%%%%%%%%%%%%%%%%%%%%%%%%%%%%%%%
%
%
%                           Standard Abbreviations
%

\def\deg{{\rm deg}\hskip 1pt}

%\hskip 1pt}
\def\SU{{\rm SU}}%\hskip 1pt}
%\hskip 1pt}

\def\End{{\rm End}\hskip 1pt}
\def\Fun{{\rm Fun}\hskip 1pt}

\def\hst1{\hskip 1pt}

%
%
%%%%%%%%%%%%%%%%%%%%%%%%%%%%%%%%%%%%%%%%%%%%%%%%%%%%%%%%%%%%%%%%%%%%%%%%%%%%%%
%
%
%

\def\titlebf#1{\vskip.5cm${\underline{\hbox{\bf #1}}}$\vskip.5cm}

\def\xxx{\phantom{xxxxxxxxxxxxxxxxxxxxxxxxxxxxx}}

%%%%%%%%%%%%%%%%%%%%%%%%%%%%%%%%%%%%%%%%%%%%%%%%%%%%%%%%%%%%%%%%%%%%%%%%%%%%%%%%%%%%

\hrule\vskip.5cm
\hbox to 16.5 truecm{September 2004 \hfil DFUB 04--03}
\hbox to 16.5 truecm{Version 1  \hfil hep-th/0409181}
\vskip.5cm\hrule
\vskip.9cm
\centerline{\bf A SIGMA MODEL FIELD THEORETIC REALIZATION}   
\centerline{\bf OF HITCHIN'S GENERALIZED COMPLEX GEOMETRY}   
\vskip.4cm
\centerline{by}
\vskip.4cm
\centerline{\bf Roberto Zucchini}
\centerline{\it Dipartimento di Fisica, Universit\`a degli Studi di Bologna}
\centerline{\it V. Irnerio 46, I-40126 Bologna, Italy}
\centerline{\it I.N.F.N., sezione di Bologna, Italy}
\centerline{\it E--mail: zucchinir@bo.infn.it}
\vskip.9cm
\hrule
\vskip.6cm
\centerline{\bf Abstract} 
\vskip.4cm
\par\noindent
We present a sigma model field theoretic realization of Hitchin's generalized 
complex geometry, which recently has been shown to be relevant in compactifications 
of superstring theory with fluxes. 
Hitchin sigma model is closely related to the well known
Poisson sigma model, of which it has the same field content.
The construction shows a remarkable correspondence between the (twisted) 
integrability conditions of generalized almost complex structures and 
the restrictions on target space geometry implied by the Batalin--Vilkovisky
classical master equation. Further, the (twisted) classical Batalin--Vilkovisky 
cohomology is related non trivially to a generalized Dolbeault cohomology. 

\vskip.6cm
\hrule
\vskip.6cm
\par\noindent
MSC-class: 53D17, 53B50. Keywords: Poisson Sigma Model, Generalized Complex Geometry,
Cohomology.
\vfill\eject

\titlebf{Contents}   

\item{1.} Introduction

\item{2.} Generalized complex geometry

\item{3.} 2--dimensional de Rham superfields

\item{4.} The Hitchin sigma model

\item{5.} The twisted Hitchin sigma model

\item{6.} Batalin--Vilkovisky cohomology and generalized complex geometry

\vfill\eject

\titlebf{1. Introduction}

\par
Mirror symmetry is a duality relating compactifications of type IIA and type IIB 
superstring theory, which result in the same four--dimensional effective theory. 
For Calabi--Yau compactifications, it has been known for a long time and it has played 
an important role in their study. Recently, more general compactifications allowing 
for non Ricci--flat metrics and NS and RR fluxes have become object of intense 
scrutiny. Therefore, it is important to investigate whether mirror symmetry 
generalizes to this more general class of compactifications and, if so, to analyze 
in depth its properties. This program was outlined originally in refs. \ref{1,2} and 
was subsequently pursued with an increasing level of generality in a series of papers. 

In refs. \ref{3}, it was shown that mirror symmetry can be defined on manifolds 
with $\SU(3)$ structure, i.e. admitting a nowhere vanishing globally defined 
internal spinor. In this case, the symmetry maps RR into RR fluxes, but it 
mixes the metric and the NS flux in a non trivial fashion. 

Recently, Hitchin formulated the notion of generalized complex geometry, which,
at the same time extends and unifies the customary notions of complex and 
symplectic geometry and incorporates a natural generalization of Calabi--Yau 
geometry \ref{4}. Hitchin's ideas were further developed by Gualtieri \ref{5}. 
Since, in topological string theory, mirror symmetry relates 
complex and symplectic manifolds \ref{6}, it is conceivable that generalized complex 
geometry may provide a natural framework for the study of mirror symmetry
\ref{7}. In refs. \ref{8,9}, it was shown that supersymmetric $\SU(3)$
structure manifolds are indeed generalized Calabi--Yau manifolds as defined by
Hitchin. Other studies of mirror symmetry relying on generalized complex geometry
can be found in refs. \ref{10--14}.

In refs. \ref{15,16}, a sigma model realization of Hitchin's generalized complex 
geometry closely resembling a Poisson sigma model was obtained \ref{17,18}. 
In this paper, we obtain a new sigma model realization of the same geometry, 
whose relation to the standard Poisson sigma model is even closer and which 
we now briefly outline. 

In ref. \ref{19} (see also \ref{20}), Cattaneo and Felder quantized the Poisson 
sigma model by using the Batalin--Vilkovisky quantization algorithm \ref{21--23}. 
They showed in particular that the action of the model satisfies the 
Batalin--Vilkovisky classical master equation, provided the target space 
almost Poisson structure is actually Poisson, thus establishing a remarkable 
connection between Poisson geometry and quantization \`a la Batalin--Vilkovisky 
of the sigma model.

In this paper, we introduce a Hitchin sigma model, which has the same field content as
the standard Poisson sigma model, but whose target space geometry is specified by a 
generalized almost complex structure. Proceeding in an analogous manner, we quantize 
the model following the Batalin--Vilkovisky quantization prescriptions. We then 
show that the action satisfies the Batalin--Vilkovisky classical master equation, when 
the generalized almost complex structure is actually a generalized complex structure. 
We carry out our analysis both in the twisted and in the untwisted case. 
Further, we find that the classical Batalin--Vilkovisky cohomology
is related non trivially to a hitherto unknown generalized Dolbeault cohomology
containing the deformation cohomology of generalized complex structures \ref{5}.

Up to a topological term, the Hitchin sigma model reduces to the usual Poisson 
sigma model, in the particular case where the generalized complex structure 
is actually a symplectic structure. In this way, our analysis partially generalizes 
and broadens the scope of Cattaneo's and Felder's. 

This paper is organized as follows.
In sect. 2, we review the main notions of generalized complex geometry 
both in the twisted and in the untwisted case. In sect. 3, we outline the 
de Rham superfield formalism suitable for the formulation of the Poisson sigma model
and its generalizations. In sect. 4, we introduce the untwisted Hitchin sigma model
and show the correspondence between the conditions on the target space geometry 
implied by the Batalin--Vilkovisky classical master equation and the integrability 
condition of the target space generalized almost complex structure. 
In sect. 5, we repeat the same analysis for the twisted case. 
Finally, in sect. 6, we analyze the Batalin--Vilkovisky cohomology
and its relation to generalized complex geometry.

\titlebf{2. Generalized complex geometry}

\par
The notion of generalized complex structure was introduced by Hitchin 
in \ref{4} and developed by Gualtieri \ref{5} in his thesis. 
It encompasses the usual notions of complex and symplectic 
structures as special cases.
It is the complex counterpart of the notion of Dirac structure, introduced by 
Courant and Weinstein, which unifies Poisson and symplectic geometry \ref{24,25}. 

Let $M$ be a manifold of even dimension $d$. Consider the vector bundle 
$TM\oplus T^*M$. A generic section $X+\xi\in C^\infty(TM\oplus T^*M)$ 
of this bundle is the direct sum of sections $X\in C^\infty(TM)$, 
$\xi\in C^\infty(T^*M)$ of $TM$, $T^*M$, respectively. 
$X$ is a vector field, $\xi$ is a $1$--form. 
 
$TM\oplus T^* M$ is equipped with a natural indefinite metric of signature  
$(d,d)$ defined by 
$$
\langle X+\xi,Y+\eta\rangle=\hbox{$1\over 2$}(i_X\eta+i_Y\xi),
\eqno(2.1)
$$
for $X+\xi, Y+\eta\in C^\infty(TM\oplus T^* M)$, where $i_V$ 
denotes contraction with respect a vector field $V$. 
This metric has a large isometry group. This contains the full 
diffeomorphism group of $M$, acting by pull-back. It also contains
the following distinguished isometries, called $b$ transforms, 
defined by 
$$
\exp(b)(X+\xi)= X+\xi+i_X b,
\eqno(2.2)
$$
where $b\in C^\infty(\wedge^2 T^* M)$ is a $2$--form. 

There is a natural bilinear pairing defined on $C^\infty(TM\oplus T^*M)$
extending the customary Lie pairing on $C^\infty(TM)$, called Courant brackets 
\ref{24,25}. It is given by 
$$
[X+\xi,Y+\eta]=[X,Y]+l_X\eta-l_Y\xi-\hbox{$1\over 2$}d_M(i_X\eta-i_Y\xi).
\eqno(2.3)
$$
with $X+\xi, Y+\eta\in C^\infty(TM\oplus T^* M)$, where $l_V$ 
denotes Lie derivation with respect a vector field $V$ and $d_M$ is 
the exterior differential of $M$. 
The pairing is antisymmetric, but it fails to satisfy the 
Jacobi identity. However, remarkably, the Jacobi identity is satisfied 
when restricting the sections $X+\xi, Y+\eta\in C^\infty(L)$,
where $L$ is a subbundle of $TM\oplus T^* M$ isotropic with respect 
to $\langle\,,\rangle$ and involutive (closed) under $[\,,]$.
The brackets $[\,,]$ are covariant under the action of the diffeomorphism group.
They are also covariant under $b$ transform
$$
[\exp(b)(X+\xi),\exp(b)(Y+\eta)]=\exp(b)[X+\xi,Y+\eta],
\eqno(2.4)
$$
provided the $2$--form $b$ is closed.

A generalized complex structure $\cal J$ is a section of 
$C^\infty(\End(TM\oplus T^* M))$, which is an isometry of 
the metric $\langle\,,\rangle$ and satisfies 
$$
{\cal J}^2=-1.
\eqno(2.5)
$$ 
The group of isometries of $\langle\,,\rangle$ acts on $\cal J$ 
by conjugation. In particular, the $b$ transform of $\cal J$ 
is defined by \xxx
$$
\hat{\cal J}=\exp(-b){\cal J}\exp(b).
\eqno(2.6)
$$ 

The $\pm\sqrt{-1}$ eigenbundles of $\cal J$ are complex and, thus, 
their analysis requires complexifying $TM\oplus T^* M$ leading to 
$(TM\oplus T^* M)\otimes\Bbb C$. The projectors on the eigenbundles are
given by \xxx
$$
\Pi_{\pm}=\hbox{$1\over 2$}(1\mp\sqrt{-1}{\cal J}).
\eqno(2.7)
$$
The generalized almost complex structure $\cal J$ is integrable if 
its eigenbundles are involutive, i. e. if \xxx
$$
\Pi_{\mp}[\Pi_{\pm}(X+\xi),\Pi_{\pm}(Y+\eta)]=0,
\eqno(2.8)
$$
for any $(X+\xi),(Y+\eta)\in C^\infty(TM\oplus T^* M)$.
In that case, $\cal J$ is called a generalized complex structure. 
Integrability is equivalent to the single statement
$$
N(X+\xi,Y+\eta)=0,
\eqno(2.9)
$$
for all $X+\xi, Y+\eta\in C^\infty(TM\oplus T^* M)$, 
where $N$ is the generalized Nijenhuis tensor, defined by 
$$
\eqalignno{\vphantom{1\over 2} 
N(X+\xi,Y+\eta)&=[X+\xi,Y+\eta]+{\cal J}[{\cal J}(X+\xi),Y+\eta]
+{\cal J}[X+\xi, {\cal J}(Y+\eta)]~~~~~~~&(2.10)\cr
\vphantom{1\over 2}
&\,-[{\cal J}(X+\xi),{\cal J}(Y+\eta)].&\cr
}
$$
The $b$ transform $\hat{\cal J}$ of a generalized complex structure $\cal J$ is 
again a generalized complex structure, provided the $2$--form $b$ is closed. 

In practice, it is convenient to decompose a generalized almost complex structure 
$\cal J$ in block form as follows \xxx
$$
 {\cal J} = \left(\matrix{J& P&\cr
                          Q& -J^*&\cr}\!\!\!\!\!\!\!\right),
\eqno(2.11)
$$
where $J\in C^\infty(TM\otimes T^* M)$, $P\in C^\infty(\wedge^2 TM)$, 
$Q\in C^\infty(\wedge^2 T^* M)$.

For later use, we write in explicit tensor notation the conditions obeyed by 
$J$, $P$, $Q$: 
$$
\eqalignno{
\vphantom{1\over 2}
&P^{ab}+P^{ba}=0,&(2.12a)\cr
\vphantom{1\over 2}
&Q_{ab}+Q_{ba}=0,&(2.12b)\cr
\vphantom{1\over 2}
&J^a{}_cJ^c{}_b+P^{ac}Q_{cb}+\delta^a{}_b=0,&(2.13a)\cr
\vphantom{1\over 2}
&J^a{}_cP^{cb}+J^b{}_cP^{ca}=0,&(2.13b)\cr
\vphantom{1\over 2}
&Q_{ac}J^c{}_b+Q_{bc}J^c{}_a=0.&(2.13c)\cr
}
$$
Under $b$ transform, we have
$$
\eqalignno{\vphantom{1\over 2} 
&\hat J^a{}_b=J^a{}_b-P^{ac}b_{cb},&(2.14a)\cr
\vphantom{1\over 2} 
&\hat P^{ab}=P^{ab},&(2.14b)\cr
\vphantom{1\over 2} 
&\hat Q_{ab}=Q_{ab}+b_{ac}J^c{}_b-b_{bc}J^c{}_a+P^{cd}b_{ca}b_{db}.&(2.14c)\cr
}
$$
where $b_{ab}+b_{ba}=0$.

The integrability condition of a generalized almost complex structure $\cal J$ 
can be cast in the form of a set of four tensorial equations 
$$
\eqalignno{
\vphantom{1\over 2} 
&A^{abc}=0,&(2.15a)\cr
\vphantom{1\over 2} 
&B_a{}^{bc}=0,&(2.15b)\cr
\vphantom{1\over 2} 
&C_{ab}{}^c=0,&(2.15c)\cr
\vphantom{1\over 2}
&D_{abc}=0,&(2.15d)\cr
}
$$
where $A$, $B$, $C$, $D$ are tensors defined by  
$$
\eqalignno{\vphantom{1\over 2} 
A^{abc}&=P^{ad}\partial_dP^{bc}+P^{bd}\partial_dP^{ca}+P^{cd}\partial_dP^{ab},
&(2.16a)\cr
\vphantom{1\over 2}
B_a{}^{bc}&=J^d{}_a\partial_dP^{bc}
+P^{bd}(\partial_aJ^c{}_d-\partial_d J^c{}_a)
-P^{cd}(\partial_aJ^b{}_d -\partial_dJ^b{}_a)
-\partial_a(J^b{}_dP^{dc}),~~~~~~~&(2.16b)\cr
\vphantom{1\over 2}
C_{ab}{}^c&=J^d{}_a\partial_dJ^c{}_b-J^d{}_b\partial_dJ^c{}_a
-J^c{}_d\partial_aJ^d{}_b+J^c{}_d\partial_bJ^d{}_a&(2.16c)\cr
\vphantom{1\over 2}
&\,+P^{cd}(\partial_dQ_{ab}+\partial_aQ_{bd}+\partial_bQ_{da}),&\cr
\vphantom{1\over 2}
D_{abc}&=J^d{}_a(\partial_dQ_{bc}+\partial_bQ_{cd}+\partial_cQ_{db})
+J^d{}_b(\partial_dQ_{ca}+\partial_cQ_{ad}+\partial_aQ_{dc})&(2.16d)\cr
\vphantom{1\over 2}
&+J^d{}_c(\partial_dQ_{ab}+\partial_aQ_{bd}+\partial_bQ_{da})
-\partial_a(Q_{bd}J^d{}_c)-\partial_b(Q_{cd}J^d{}_a)-\partial_c(Q_{ad}J^d{}_b).&\cr
}
$$
The above expressions in a different but equivalent form were derived in \ref{16}. 

The usual complex structures $J$ can be viewed as generalized complex structures 
of the special form \xxx
$$
 {\cal J} = \left (\matrix{J & 0 &\cr
                           0 & -J^*&\cr}\!\!\!\!\!\!\right).
\eqno(2.17)
$$
Indeed, one can check an object of this form satisfies conditions (2.12$a$,$b$), 
(2.13$a$--$c$), (2.15$a$--$d$) precisely when $J$ is a complex structure, i. e. its 
Nijenhuis tensor vanishes. 
Similarly, the usual symplectic structures $Q$ can be viewed as 
generalized complex structures of the special form \xxx
$$
{\cal J} = \left (\matrix{0 & -Q^{-1} &\cr
                          Q & 0&\cr}\!\!\!\!\!\!\right).
\eqno(2.18)
$$
as this object satisfies (2.12$a$,$b$), (2.13$a$--$c$), 
(2.15$a$--$d$) precisely when $Q$ is a 
symplectic structure, i. e. it is closed. As noticed by Hitchin, other exotic 
examples exist. In fact, there are manifolds which cannot support any complex or 
symplectic structure, but do admit generalized complex structures \ref{4}. 
These facts explain the reason why Hitchin's construction is interesting 
and worthwhile pursuing.

Let $H\in C^\infty(\wedge^3 T^*M)$ be a closed $3$--form.
Define the $H$ twisted Courant brackets by
$$
[X+\xi,Y+\eta]_H=[X+\xi,Y+\eta]+i_Xi_Y H
\eqno(2.19)
$$
with $X+\xi, Y+\eta\in C^\infty(TM\oplus T^* M)$.
\footnote{}{}\footnote{${}^1$}{The sign convention of the $H$ field
used in this paper is opposite to that of ref. \ref{5}.} 
Under $b$ transform with $b$ a closed $2$--form, (2.4) holds with the brackets
$[\,,]$ replaced by $[\,,]_H$. More generally, for a non closed $b$, one has
$$
[\exp(b)(X+\xi),\exp(b)(Y+\eta)]_{H+d_Mb}=\exp(b)[X+\xi,Y+\eta]_H.
\eqno(2.20)
$$
So, $b$ transform shifts $H$ by the exact $3$--form $d_Mb$:
$$
\hat H=H+d_Mb.
\eqno(2.21)
$$
The case where $[H/2\pi]$ belongs to the image of $H^3(M,\Bbb Z)$
in $H^3(M,\Bbb R)$ is particular important for its relation to gerbes. 
In this case, $b$ transform with $[b/2\pi]$ contained in image of $H^2(M,\Bbb Z)$
in $H^2(M,\Bbb R)$ represents the gerbe generalization of gauge transformation.

One can define an $H$ twisted generalized Nijenhuis $N_H$ tensor as in
(2.10) by using the brackets $[\,,]_H$ instead of $[\,,]$.
A generalized almost complex structure $\cal J$ is $H$ integrable if \xxx
$$
N_H(X+\xi,Y+\eta)=0,
\eqno(2.22)
$$
for all $X+\xi,Y+\eta\in C^\infty(TM\oplus T^* M)$.
In such a case, we call $\cal J$ an $H$ twisted generalized complex structure. 

In tensor notation,  the $H$ integrability conditions can be cast as 
$$
\eqalignno{\vphantom{1\over 2} 
&A_H{}^{abc}=0,&(2.23a)\cr
\vphantom{1\over 2} 
&B_H{}_a{}^{bc}=0,&(2.23b)\cr
\vphantom{1\over 2} 
&C_H{}_{ab}{}^c=0,&(2.23c)\cr
\vphantom{1\over 2}
&D_H{}_{abc}=0,&(2.23d)\cr
}
$$
where $A_H$, $B_H$ $C_H$, $D_H$ are tensors defined by  $$
\eqalignno{\vphantom{1\over 2} 
&A_H{}^{abc}=A^{abc},&(2.24a)\cr
\vphantom{1\over 2}
&B_H{}_a{}^{bc}=B_a{}^{bc}+P^{bd}P^{ce}H_{ade},&(2.24b)\cr
\vphantom{1\over 2}
&C_H{}_{ab}{}^c=C_{ab}{}^c-J^d{}_aP^{ce}H_{bde}+J^d{}_bP^{ce}H_{ade},&(2.24c)\cr
\vphantom{1\over 2}
&D_H{}_{abc}=D_{abc}-H_{abc}+J^d{}_aJ^e{}_bH_{cde}+J^d{}_bJ^e{}_cH_{ade}
+J^d{}_cJ^e{}_aH_{bde}. &(2.24d)\cr
}
$$
These expressions also where obtained in \ref{16}.

\titlebf{3. 2--dimensional de Rham superfields}

\par
In general, the fields of a 2--dimensional field theory are differential 
forms on a oriented closed $2$--dimensional manifold $\Sigma$. They can be viewed 
as elements of the space $\Fun(\Pi T\Sigma)$ of functions on the parity 
reversed tangent bundle $\Pi T\Sigma$ of $\Sigma$, which we shall call 
de Rham superfields \ref{19}. 
More explicitly, we associate with the coordinates $z^\alpha$ of 
$\Sigma$ Grassmann odd partners $\zeta^\alpha$ with \xxx
$$
\deg z^\alpha=0, \qquad \deg\zeta^\alpha=1.\vphantom{\Big[}
\eqno(3.1)
$$
$\Pi T\Sigma$ is endowed with a natural differential $d$ defined by 
$$
dz^\alpha=\zeta^\alpha,\qquad d\zeta^\alpha=0.\vphantom{\Big[}
\eqno(3.2)
$$
A generic de Rham superfield $\psi(z,\zeta)$ is a triplet 
formed by a $0$--, $1$--, $2$--form field $\psi^{(0)}(z)$, 
$\psi^{(1)}{}_\alpha(z)$, $\psi^{(2)}{}_{\alpha\beta}(z)$ 
organized as
$$
\psi(z,\zeta)=\psi^{(0)}(z)+\zeta^\alpha\psi^{(1)}{}_\alpha(z)
+\hbox{$1\over 2$}\zeta^\alpha\zeta^\beta\psi^{(2)}{}_{\alpha\beta}(z).
\eqno(3.3)
$$
The forms $\psi^{(0)}$, $\psi^{(1)}$, $\psi^{(2)}$ are called the components 
of $\psi$. Note that, in this formalism, the exterior differential of $\Sigma$ 
can be identified with the operator 
$$
d=\zeta^\alpha\partial/\partial z^\alpha.
\eqno(3.4)
$$

The coordinate invariant integration measure of $\Pi T\Sigma$ is 
$$
\mu={\rm d}z^1{\rm d}z^2{\rm d}\zeta^1{\rm d}\zeta^2.
\eqno(3.5)
$$
Any de Rham superfield $\psi$ can be integrated on $\Pi T\Sigma$ according to
the prescription
$$
\int_{\Pi T\Sigma}\mu\,\psi=\int_\Sigma\hbox{$1\over 2$}
dz^\alpha dz^\beta\psi^{(2)}{}_{\alpha\beta}(z).
\eqno(3.6)
$$
By Stokes' theorem, \xxx 
$$
\int_{\Pi T\Sigma}\mu\, d\psi=0.
\eqno(3.7)
$$

It is possible to define functional derivatives of functionals of de Rham superfields.
Let $\psi$ be a de Rham superfield and let $F(\psi)$ be a functional of $\psi$.
We define the left/right functional derivative superfields 
$\delta_{l,r} F(\psi)/\delta\psi$ as follows. 
Let $\sigma$ be a superfield of the same properties as $\psi$.
Then, 
$$
{d\over dt}F(\psi+t\sigma)\Big|_{t=0}
=\int_{\Pi T\Sigma}\mu\,\sigma{\delta_l F(\psi)\over\delta\psi}
=\int_{\Pi T\Sigma}\mu\,{\delta_r F(\psi)\over\delta\psi}\sigma.
\eqno(3.8)
$$

In the applications below, the components of the relevant de Rham superfields
carry, besides the form degree, also a ghost degree. We shall limit ourselves 
to homogeneous superfields. A de Rham superfield $\psi$ is said homogeneous 
if the sum of the form and ghost degree is the same for all its components 
$\psi^{(0)}$, $\psi^{(1)}$, $\psi^{(2)}$ of $\psi$. 
The common value of that sum is called the (total) degree $\deg\psi$ of $\psi$. 
It is easy to see that the differential operator $d$ and the
integration operator $\int_{\Pi T\Sigma}\mu$ carry 
degree $1$ and $-2$, respectively. 
Also, if $F(\psi)$ is a functional of a superfield $\psi$, then
$\deg \delta_{l,r} F(\psi)/\delta\psi=\deg F-\deg \psi+2$.

The singular chain complex of $\Sigma$ can be given a parallel treatment. 
A singular superchain $C$ is a triplet formed by a $0$--, 
$1$-- and $2$--dimensional singular chain $C_{(0)}$, $C_{(1)}$, $C_{(2)}$
organized as a formal chain sum \xxx
$$
C=C_{(0)}+C_{(1)}+C_{(2)}.
\eqno(3.9)
$$
The singular boundary operator $\partial$ extends to superchains in obvious 
fashion by setting
$$
(\partial C)_{(0)}=\partial C_{(1)}, 
\quad (\partial C)_{(1)}=\partial C_{(2)},
\quad (\partial C)_{(2)}=0.
\eqno(3.10)
$$
A singular supercycle $Z$ is a superchain such that 
$$
\partial Z=0.
\eqno(3.11)
$$

A de Rham superfield $\psi$ can be integrated on a superchain $C$:
$$
\int_C\psi=\int_{C_{(0)}}\psi^{(0)}
+\int_{C_{(1)}}dz^\alpha\psi^{(1)}{}_\alpha(z)
+\int_{C_{(2)}}\hbox{$1\over 2$}
dz^\alpha dz^\beta\psi^{(2)}{}_{\alpha\beta}(z).
\eqno(3.12)
$$
Stokes' theorem states that \xxx
$$
\int_C d\psi=\int_{\partial C} \psi.
\eqno(3.13)
$$
In particular, \xxx
$$
\int_Zd\psi=0,
\eqno(3.14)
$$
if $Z$ is a supercycle.

\titlebf{4. The Hitchin sigma model}

In this section, we shall first briefly review the formulation of the standard Poisson
sigma model based on the Batalin--Vilkovisky quantization scheme \ref{21,22}
worked out by Cattaneo and Felder in \ref{19} (see also \ref{20,26}).
To make the treatment as simple and transparent as possible, 
we shall use the convenient de Rham superfield formalism
outlined above. Expressions in terms of components are straightforward to obtain, 
though they are rather lengthy and unwieldy.
Subsequently, we introduce the Hitchin sigma model 
as a closely related partial generalization of the former. 
We shall limit ourselves to the 
lowest order in perturbation theory, since this constraints on target space geometry 
following from the Batalin--Vilkovisky classical master equation lead directly to 
Hitchin's generalized complex geometry. Quantum corrections will presumably yield
a deformation of the latter, whose study is beyond the scope of this paper. 
We will not attempt the gauge fixing of the field theory, which, at any rate, 
is expected to be essentially identical to that of the ordinary Poisson sigma model
as described in \ref{19,20}. For clarity, we shall treat first the untwisted case. 

The basic fields of the standard Poisson sigma model are a degree $0$
superembedding $x\in\Gamma(\Pi T\Sigma, M)$ and a degree $1$
supersection $y\in\Gamma(\Pi T\Sigma, x^*\Pi T^*M)$. 
With respect to each local coordinate $t^a$ of $M$, 
$x$, $y$ are given as de Rham superfields $x^a$, $y_a$. 
Under a change of coordinates, these transform as \xxx
$$
x'^a=t'^a\circ t^{-1}(x),
\eqno(4.1)
$$
$$
y'_a={\partial t^b\over\partial t'^a}\circ t^{-1}(x)
y_b.
\eqno(4.2)
$$
The resulting transformation rules of the de Rham components of $x^a(z,\zeta)$, 
$y_a(z,\zeta)$ are obtainable by expanding these relations in powers of 
$\zeta^\alpha$.

We identify the fields and antifields with $x^a$ and $y_b$, respectively.  
The Batalin--Vilkovisky odd symplectic form is 
$$
\Omega_{BV}=\int_{\Pi T\Sigma}\mu\,\delta x^a\delta y_a.
\eqno(4.3)
$$
Therefore, the Batalin--Vilkovisky antibrackets are given by
$$
(F,G)=\int_{\Pi T\Sigma}\mu\,\bigg[
{\delta_r F\over\delta x^a}{\delta_l G\over\delta y_a}-
{\delta_r F\over\delta y_a}{\delta_l G\over\delta x^a}\bigg],
\eqno(4.4)
$$
for any two functionals $F$, $G$ of $x^a$, $y_a$. 

The target space geometry of the standard Poisson sigma model is specified by 
an almost Poisson structure, that is a $2$--vector $P$. 
The action of the model is 
$$
S=\int_{\Pi T\Sigma}\mu\,\Big[y_adx^a+\hbox{$1\over 2$}P^{ab}(x)y_ay_b\Big].
\eqno(4.5)
$$

The consistent quantization of the model requires tha $S$ satisfies the 
classical Batalin--Vilkovisky master equation \xxx
$$
(S,S)=0.
\eqno(4.6)
$$ 
By a straightforward computation one finds 
$$
(S,S)=2\int_{\Pi T\Sigma}\mu\, 
\Big[-\hbox{$1\over 6$}A^{abc}(x)y_ay_by_c\Big],
\eqno(4.7)
$$
where $A$ is given by (2.16$a$). 
Hence, $S$ satisfies (4.6), if (2.15$a$a) holds. As is well--known,
condition (2.15$a$) ensures the almost Poisson structure $P$ is 
actually Poisson, so that $M$ is a Poisson manifold \ref{27}. 

The Batalin--Vilkovisky variations are
$$
\eqalignno{\vphantom{1\over 2} 
\delta_{BV} x^a&=(S,x^a),&(4.8a)\cr
\vphantom{1\over 2}
\delta_{BV} y_a&=(S,y_a)&(4.8b)\cr
}
$$
\ref{21,22}. From (4.4), (4.5), one finds easily that \xxx
$$
\eqalignno{\vphantom{1\over 2} 
\delta_{BV} x^a&=dx^a+P^{ab}(x)y_b,&(4.9a)\cr
\vphantom{1\over 2}
\delta_{BV} y_a&=dy_a+\hbox{$1\over 2$}\partial_aP^{bc}(x)y_by_c.
&(4.9b)\cr
}
$$
$\delta_{BV}$ is nilpotent: \xxx
$$
\delta_{BV}{}^2=0
\eqno(4.10)
$$
if (2.15a) holds, as follows from the Batalin--Vilkovisky theory or by direct 
verification. By construction, \xxx
$$
\delta_{BV} S=0.
\eqno(4.11)
$$

We construct the Hitchin sigma model as follows. The fields of the model, the 
Batalin--Vilkovisky odd symplectic form and the associated antibrackets are the 
same as those of the Poisson sigma model given by  eqs. (4.3), (4.4). The target 
space geometry is specified by a generalized almost complex structure 
$\cal J$ (cf. sect. 2). In the representation (2.11), the action of the model reads 
$$
S=\int_{\Pi T\Sigma}\mu\,\Big[y_adx^a+\hbox{$1\over 2$}P^{ab}(x)y_ay_b
+\hbox{$1\over 2$}Q_{ab}(x)dx^adx^b+J^a{}_b(x)y_adx^b\Big].
\eqno(4.12)
$$

We now verify under which conditions $S$ satisfies the classical Batalin--Vilkovisky 
master equation (4.6). By a straightforward computation, one finds
$$
\eqalignno{\vphantom{1\over 2} 
(S,S)=
2\int_{\Pi T\Sigma}\mu\,\Big[&-\hbox{$1\over 6$}A^{abc}(x)y_ay_by_c
+\hbox{$1\over 2$}B_a{}^{bc}(x)dx^ay_by_c&(4.13)\cr
\vphantom{1\over 2}
&-\hbox{$1\over 2$}C_{ab}{}^c(x)dx^adx^by_c
+\hbox{$1\over 6$}D_{abc}(x)dx^adx^bdx^c\Big],
&\cr}
$$
where the tensors $A$, $B$, $C$, $D$ are given by (2.16$a$--$d$). 
Hence, $S$ satisfies the classical Batalin--Vilkovisky master equation
(4.6), if conditions (2.15$a$--$d$) hold. (2.15$a$--$d$) ensure that $\cal J$ 
is a generalized complex structure so that $M$ is a generalized complex manifold.
This shows that {\it there is a non trivial connection between generalized complex 
geometry and quantization \`a la Batalin--Vilkovisky of the sigma model.}

(2.15$a$--$d$) are sufficient but not necessary conditions for the fulfillment 
of the master equation (4.6). In fact, as $dx^adx^bdx^c=0$ identically on a 
$2$--dimensional manifold $\Sigma$, the last term in (4.13) vanishes identically
so that condition (2.15$d$) could be dropped. 
Remarkably, however, a formal implementation of the antibracket 
algebra yields precisely this term. At any rate, this opens the possibility that 
the Hitchin sigma model might consistently be defined for a class of target space 
geometries wider than that of generalized complex structures. This would somewhat 
parallel what happens for the Poisson sigma model, which makes sense for the class of 
Poisson target space geometries, which strictly contains that of symplectic 
geometries. We will not elaborate further on this point. 

The Batalin--Vilkovisky variations, defined by (4.8$a$,$b$), can be easily 
obtained using (4.4), (4.12). They are given by 
$$
\eqalignno{\vphantom{1\over 2} 
\delta_{BV} x^a&=dx^a+P^{ab}(x)y_b+J^a{}_b(x)dx^b,&(4.14a)\cr
\vphantom{1\over 2}
\delta_{BV} y_a&=dy_a+\hbox{$1\over 2$}\partial_aP^{bc}(x)y_by_c
+\hbox{$1\over 2$}(\partial_aQ_{bc}+\partial_bQ_{ca}+\partial_cQ_{ab})(x)dx^bdx^c&
(4.14b)\cr
\vphantom{1\over 2}
&+(\partial_aJ^b{}_c-\partial_cJ^b{}_a)(x)y_bdx^c+J^b{}_a(x)dy_b.&\cr
}
$$
One can check that (4.10), (4.11) hold, if eqs. (2.15$a$--$d$) are fulfilled. 

When the generalized complex structure $\cal J$ is a symplectic structure $Q$, 
one has $dQ=0$, $J=0$, $P=-Q^{-1}$ (cf. eq. (2.18)). In this case, as is readily 
verified, the Hitchin sigma 
model action equals the Poisson sigma model action up to a topological term, while the
Batalin--Vilkovisky variations of the two models are the same. 
For this reason, 
Hitchin sigma model is only a partial generalization of the Poisson sigma 
model, since it can reproduce the latter only in the particular case where the target 
manifold Poisson structure is symplectic.

It is interesting to see how the action $S$ behaves under a $b$ transform 
of the underlying generalized almost complex structure $\cal J$
of the form (2.14$a$--$c$). 
It turns out that a meaningful comparison of the resulting action $\hat S$ and the 
original action $S$ requires that the superfields $x^a$, $y_a$ also must undergo a 
$b$ transform of the form 
$$
\eqalignno{\vphantom{1\over 2} 
&\hat x^a=x^a, &(4.15a)\cr
\vphantom{1\over 2}
&\hat y_a=y_a+b_{ab}(x)dx^b.&(4.15b)\cr
}
$$
It is simple to verify that 
$$
\hat\Omega_{BV}=\Omega_{BV}+\int_{\Pi T\Sigma}\mu\,\hbox{$1\over 2$}
\big(\partial_ab_{bc}+\partial_bb_{ca}+\partial_cb_{ab}\big)(x)
\delta x^a dx^b\delta x^c. 
\eqno(4.16)
$$ 
Hence, when the $2$--form $b$ is closed, the $b$ transform is canonical, i. e. 
it leaves the Batalin--Vilkovisky odd symplectic form (4.3) invariant,
$$
\hat\Omega_{BV}=\Omega_{BV}.
\eqno(4.17)
$$
Under the $b$ transform, one has 
$$
\hat S=S-\int_{\Pi T\Sigma}\mu\, b_{ab}(x)dx^adx^b.
\eqno(4.18)
$$
Remarkably, $S$, $\hat S$ differ by a topological term. If $b/2\pi$ 
has integer periods and, so, describes a gerbe gauge transformation,
one has $\exp(\sqrt{-1}\hat S)=\exp(\sqrt{-1}S)$ in the quantum path integral. 
So, {\it gerbe gauge transformation is a duality symmetry of the quantum 
Hitchin sigma model.} It must be remarked that the result holds also when $b/2\pi$ 
has half integer periods. This is due to a factor $1/2$ mismatch of the 
normalization of the $b$ field in Hitchin's formulation of generalized 
complex geometry and the standard conventional normalization of the Poisson 
sigma model action. We would suggest to change the normalization of the action
by a factor $1/2$, but, for the time being, we stick to the normalization 
conventions mostly used in the physical literature.

\titlebf{5. The twisted Hitchin sigma model}

The natural question arises whether it is possible to construct a Hitchin sigma 
model for twisted generalized complex structures. We are going to study this issue
next. 

Let $H$ be a closed $3$--form. Consider a $H$ twisted generalized complex 
structure $\cal J$. Then, the tensors $A_H$, $B_H$, $C_H$, $D_H$
all vanish (cf. eqs. (2.23$a$--$d$)). 

Since $H$ is closed, it can be trivialized locally. So, there
are locally defined  $2$--forms $B$ such that \xxx
$$
H=d_MB.
\eqno(5.1)
$$
We can use the local $2$--forms $B$ to carry out local $B$ transforms of $\cal J$.
In this way, for each local $2$--form $B$, we have a local generalized almost complex
structure $\tilde{\cal J}$ given by 
$$
\eqalignno{\vphantom{1\over 2} 
&\tilde J^a{}_b=J^a{}_b-P^{ac}B_{cb},&(5.2a)\cr
}
$$
$$
\eqalignno{
\vphantom{1\over 2} 
&\tilde P^{ab}=P^{ab},&(5.2b)\cr
\vphantom{1\over 2} 
&\tilde Q_{ab}=Q_{ab}+B_{ac}J^c{}_b-B_{bc}J^c{}_a+P^{cd}B_{ca}B_{db}&(5.2c)\cr
}
$$
in the representation (2.11) (cf. eqs. (2.14$a$--$c$)). 
It is straightforward to verify that the $H$ integrability 
of $\cal J$ implies that these local $\tilde{\cal J}$ are integrable: the 
corresponding local tensors $\tilde A$, $\tilde B$, $\tilde C$, $\tilde D$ 
all vanish (cf. eqs. (2.15$a$--$d$)).

Let us assume that the appropriate fields/antifields for the twisted version of 
Hitchin sigma model are the same as those of the untwisted model, 
viz the degree $0$ superembedding $x\in\Gamma(\Pi T\Sigma, M)$ and the 
degree $1$ supersection $y\in\Gamma(\Pi T\Sigma, x^*\Pi T^*M)$. 
This is reasonable, since in the limit case $H=0$, one should recover the 
untwisted model.  
In the previous section, we learned that these fields behave non trivially under 
$b$ transform (cf. eqs. (4.15$a$,$b$)). 
Thus, it seems appropriate to define $B$ transformed
fields $\tilde x^a$, $\tilde y_a$ by
$$
\eqalignno{\vphantom{1\over 2} 
&\tilde x^a=x^a, &(5.3a)\cr
\vphantom{1\over 2}
&\tilde y_a=y_a+B_{ab}(x)dx^b. &(5.3b)\cr
}
$$
Since the $2$--form $B$ is only locally defined, the superfields 
$\tilde x^a$, $\tilde y_a$ do not have the global meaning that the original 
superfields $x^a$, $y_a$ do.   

From the above discussion, it would seem that, at a heuristic level, one may define 
the twisted Hitchin sigma model with target space geometry specified by the 
data $H$, $\cal J$ and basic superfields $x$, $y$ as the untwisted Hitchin sigma model 
with target space geometry specified by the data $\tilde{\cal J}$ and basic 
superfields $\tilde x$, $\tilde y$. However, it is clear that this way of proceeding 
cannot work in general, because $\tilde{\cal J}$, $\tilde x$, $\tilde y$
have only a local nature. 

There is however a particular case where this can be done, namely when the 
closed $3$--form $H$ is exact. In that case, there is a globally defined $2$--form 
$B$ such that (5.1) holds. Then, $\tilde{\cal J}$ is a globally defined generalized
complex structure and $\tilde x$, $\tilde y$ have the same global meaning as
the original fields $x$, $y$. In this way, we can construct an untwisted Hitchin 
sigma model using $\tilde{\cal J}$, $\tilde x$, $\tilde y$, which now we describe. 
As we shall see, from this analysis, we can learn much on the $H$ twisted Hitchin 
sigma model for non exact $H$.

The odd symplectic form of the $H$ twisted model is defined by the relation
$$
\Omega_{BVH}=\tilde\Omega_{BV},
\eqno(5.4)
$$
where $\tilde\Omega_{BV}$ is given by (4.3) with $x^a$, $y_a$ replaced by 
$\tilde x^a$, $\tilde y_a$. A straightforward calculation shows that \xxx
$$
\Omega_{BVH}=\int_{\Pi T\Sigma}\mu\,\Big[\delta x^a\delta y_a
+\hbox{$1\over 2$}H_{abc}(x)\delta x^a dx^b \delta x^c\Big].
\eqno(5.5)
$$
$\Omega_{BVH}$ is not of the canonical form (4.3). Hence, $x^a$, $y_a$ 
are not canonical fields/antifields. However, $\Omega_{BVH}$ is a degree $1$
closed functional form, since it is related by a field redefinition to 
$\tilde\Omega_{BV}$, which is. In this way, one can define $H$ twisted 
antibrackets $(,)_H$ in standard fashion. The resulting expression is 
$$
(F,G)_H=\int_{\Pi T\Sigma}\mu\,\bigg[
{\delta_r F\over\delta x^a}{\delta_l G\over\delta y_a}-
{\delta_r F\over\delta y_a}{\delta_l G\over\delta x^a}
-H_{abc}(x){\delta_r F\over\delta y_a} 
dx^b {\delta_l G\over\delta y_c}\bigg],
\eqno(5.6)
$$
for any two functionals $F$, $G$ of $x^a$, $y_a$.

Similarly, the action of the $H$ twisted model is defined by the relation
$$
S_H=\tilde S, 
\eqno(5.7)
$$
where $\tilde S$ is given by (4.12) with $J$, $P$, $Q$ and 
$x^a$, $y_a$ replaced by $\tilde J$, $\tilde P$, $\tilde Q$ and 
$\tilde x^a$, $\tilde y_a$, respectively. A straightforward calculation shows that
$$
S_H=\int_{\Pi T\Sigma}\mu\,\Big[y_adx^a+\hbox{$1\over 2$}P^{ab}(x)y_ay_b
+\hbox{$1\over 2$}Q_{ab}(x)dx^adx^b+J^a{}_b(x)y_adx^b\Big]
-2\int_\Gamma x^{(0)*}H.
\eqno(5.8)
$$
Here, $\Gamma$ is a $3$--fold such that $\partial \Gamma=\Sigma$ and 
$x^{(0)}:\Gamma\rightarrow M$ is an embedding such that $x^{(0)}|_\Sigma$
equals the lowest degree $0$ component of the superembedding $x$, 
whose choice is immaterial.

It is easy to see that the $H$ twisted action $S_H$ obeys the $H$ twisted 
Batalin--Vilkovisky classical master equation \xxx 
$$
(S_H, S_H)_H=0.
\eqno(5.9)
$$
This can be verified directly, but it is obvious by itself, since 
$\Omega_{BVH}$, $S_H$ are related to $\tilde\Omega_{BV}$, $\tilde S$
by the same field redefinition, via (5.4), (5.7), respectively, and the action 
$\tilde S$ obeys the master equation (4.6). 

The Batalin--Vilkovisky variations $\tilde\delta_{BV}\tilde x^a$, 
$\tilde\delta_{BV}\tilde y_a$ are given by (4.14$a$,$b$) 
with $J$, $P$, $Q$ and $x^a$, $y_a$ replaced by $\tilde J$, 
$\tilde P$, $\tilde Q$ and $\tilde x^a$, $\tilde y_a$, respectively.
Using (5.3$a$,$b$), we can then derive the expressions of the $H$ twisted 
Batalin--Vilkovisky variations $\delta_{BVH} x^a$, $\delta_{BVH} y_a$. The result is 
$$
\eqalignno{
\vphantom{1\over 2} 
\delta_{BVH} x^a&=dx^a+P^{ab}(x)y_b+J^a{}_b(x)dx^b,&(5.10a)\cr
\vphantom{1\over 2}
\delta_{BVH} y_a&=dy_a+\hbox{$1\over 2$}\partial_aP^{bc}(x)y_by_c
+\hbox{$1\over 2$}(\partial_aQ_{bc}+\partial_bQ_{ca}+\partial_cQ_{ab})(x)dx^bdx^c&
(5.10b)\cr
\vphantom{1\over 2}
&+(\partial_aJ^b{}_c-\partial_cJ^b{}_a)(x)y_bdx^c+J^b{}_a(x)dy_b\cr
}
$$
$$
\eqalignno{
\vphantom{1\over 2}
&+\hbox{$1\over 2$}(H_{abd}J^d{}_c-H_{acd}J^d{}_b)(x)dx^bdx^c
+H_{adc}P^{db}(x)y_bdx^c.&\cr
}
$$
It is straightforward to see that \xxx
$$
\eqalignno{\vphantom{1\over 2} 
\delta_{BVH} x^a&=(S_H,x^a)_H,&(5.11a)\cr
\vphantom{1\over 2}
\delta_{BVH} y_a&=(S_H,y_a)_H &(5.11b)\cr
}
$$
and that \xxx
$$
\delta_{BVH}{}^2=0.
\eqno(5.12)
$$
Again, this can be verified directly, but it is obvious by itself, for reasons 
explained below eq. (5.9). 
By a similar reasoning, one has \xxx
$$
\delta_{BVH} S_H=0.
\eqno(5.13)
$$

In the above analysis, we assumed that the closed $3$--form $H$ was exact, in 
order to have a globally defined $2$--form $B$. However, the expressions obtained
depend on $B$ through $H$, which is globally defined anyway. This provides 
crucial clues about how to proceed for a non exact $H$.

We can use (5.5) as a definition of the $H$ twisted odd symplectic form 
$\Omega_{BVH}$ 
in the general case. $\Omega_{BVH}$ so defined has degree $1$ and is closed,
as required. So, $H$ twisted antibrackets $(,)_H$  can be introduced.
They are given again by eq. (5.6).

Similarly, we can use (5.8) as a definition of the $H$ twisted action $S_H$
in the general case. At this stage, $\cal J$ can be assumed to be
a generalized almost complex structure. The last $H$ dependent term is 
a Wess--Zumino like term. A similar term was added to the action of 
the standard Poisson sigma model in ref. \ref{28}. 
Its value depends on the embedding 
$x^{(0)}:\Gamma\rightarrow M$. In the quantum theory, in order to have a 
well defined weight $\exp(\sqrt{-1}S_H)$ in the path integral, 
it necessary to require that $H$ has integer periods.

A computation analogous to the one leading to (4.13) furnishes 
$$
\eqalignno{\vphantom{1\over 2} 
(S_H,S_H)_H=
2\int_{\Pi T\Sigma}\mu\,\Big[&-\hbox{$1\over 6$}A_H{}^{abc}(x)y_ay_by_c
+\hbox{$1\over 2$}B_H{}_a{}^{bc}(x)dx^ay_by_c&(5.14)\cr
\vphantom{1\over 2}
&-\hbox{$1\over 2$}C_H{}_{ab}{}^c(x)dx^adx^by_c
+\hbox{$1\over 6$}D_H{}_{abc}(x)dx^adx^bdx^c&\cr
\vphantom{1\over 2}
&-\hbox{$1\over 3$}H_{abc}(x)dx^adx^bdx^c\Big],
&\cr}
$$
where the tensors $A_H$, $B_H$, $C_H$, $D_H$ 
are given by (2.24$a$--$d$). Hence, $S_H$ satisfies the $H$ twisted classical 
Batalin--Vilkovisky master equation (5.9), if (2.23$a$--$d$) hold, i.e. when
$\cal J$ is an $H$ twisted generalized complex structure. (Recall 
that $dx^adx^bdx^c=0$ on a $2$--dimensional manifold $\Sigma$.)
This shows that {\it the non trivial connection between generalized complex 
geometry and quantization \`a la Batalin--Vilkovisky of the sigma model continues 
to hold also in the twisted case.} As for the untwisted Hitchin model, 
(2.24$a$--$d$) are sufficient but not necessary conditions for the fulfillment 
of the master equation (5.9). 

The twisted Batalin--Vilkovisky variations $\delta_{BVH} x^a$, $\delta_{BVH} y_a$ 
are defined by (5.11$a$,$b$). By explicit computation, one can verify that 
they are still given by (5.10$a$,$b$). This is obviously so in view of the above 
reasoning, since these are local expressions anyway. Similarly, 
(5.12), (5.13) continue to hold. 

Under a $b$ transform (2.14$a$--$c$) of the underlying 
generalized almost complex structure and (4.15$a$,$b$) of the superfields 
$x^a$, $y_a$, with $b$ a closed $2$--form, the $H$ twisted odd symplectic form 
$\Omega_{BVH}$ and action $S_H$ behave as their untwisted counterparts,
that is (4.17), (4.18) hold with $\Omega_{BV}$, $\hat\Omega_{BV}$ 
$S$, $\hat S$ replaced by $S_H$, $\hat S_H$, $\Omega_{BVH}$, $\hat\Omega_{BVH}$.

\titlebf{6. Batalin--Vilkovisky cohomology and generalized complex geometry}

In this final section we shall analyze the classical Batalin--Vilkovisky cohomology 
and its relation to Hitchin's (twisted) generalized complex geometry. Though we do 
not have a full computation of the cohomology, we have found a interesting subset 
of it related in a non trivial fashion to the underlying generalized complex 
structure.

We consider first the untwisted case for simplicity. 
We call a de Rham superfield $X$ local, if it is a local functional of the basic 
superfields $x^a$, $y_a$. Let $X$ be some local superfield. Suppose there is 
another local superfield $Y$ such that 
$$
\delta_{BV}X=dY.
\eqno(6.1)
$$
Thus, $X$ defines a mod $d$ Batalin--Vilkovisky cohomology class.
Then, if $Z$ is a singular supercycle (cf. sect. 3), one has \xxx
$$
\delta_{BV}\int_ZX=\int_ZdY=0,
\eqno(6.2)
$$
by (3.14). It follows that \xxx
$$
\langle Z, X\rangle=\int_ZX
\eqno(6.3)
$$
defines a Batalin--Vilkovisky cohomology class. A standard analysis shows that this 
class depends only on the mod $d$ Batalin--Vilkovisky cohomology class of $X$. 
%and singular homology class of $Z$.
So, one may obtain Batalin--Vilkovisky cohomology classes by constructing 
local superfields $X$ satisfying (6.1).

We recall that \xxx
$$
d\delta_{BV}+\delta_{BV}d=0.
\eqno(6.4)
$$
Define \xxx
$$
\partial=\hbox{$1\over 2$}\big[d-\sqrt{-1}(\delta_{BV}-d)\big]
\eqno(6.5)
$$
and its complex conjugate $\overline\partial$.
From (4.10), (6.4), it is immediate to check that 
$$
\eqalignno{
\vphantom{1\over 2}
\partial^2&=0,&(6.6a)\cr
\vphantom{1\over 2}
\overline\partial{}^2&=0,&(6.6b)\cr
\vphantom{1\over 2}
\partial\overline\partial+\overline\partial\partial&=0.&(6.6c)\cr
}
$$
From (6.5), one has further \xxx
$$
\eqalignno{
\vphantom{1\over 2}
d&=\partial+\overline\partial,&(6.7a)\cr
\vphantom{1\over 2}
\delta_{BV}&=\partial+\overline\partial
+\sqrt{-1}(\partial-\overline\partial).
&(6.7b)\cr
}
$$

Consider the operator $\overline\partial$. It acts on the space of local superfields, 
it carries degree $1$, by (6.5), and it squares to $0$, by (6.6$b$). Therefore, one 
can define a $\overline\partial$ local superfield cohomology in obvious fashion.

Let $X$ be a local superfield such that 
$$
\overline\partial X=0.
\eqno(6.8)
$$
$X$ defines a $\overline\partial$ local superfield cohomology class. 
By (6.7$a,b$), $X$ satisfies (6.1) with $Y=(1+\sqrt{-1})X$. So, as shown above, 
for any supercycle $Z$, $\langle Z, X\rangle$ defines a Batalin--Vilkovisky 
cohomology class. If $X=\overline \partial U$ for some local superfield $U$,
so that the corresponding $\overline\partial$ cohomology class is trivial, 
then $\langle Z, X\rangle={1\over 2}\sqrt{-1}\delta_{BV}\langle Z, U\rangle$, 
by (6.5), (3.14),  and, so, the corresponding Batalin--Vilkovisky 
is trivial as well. Therefore, for any singular supercycle $Z$,
there is a well-defined homomorphism from the $\overline\partial$ superfield 
cohomology into the Batalin--Vilkovisky cohomology. This 
homomorphism depnds only on singular homology class of $Z$.

The above constructions may appear somewhat arbitrary. Their meaningfulness will 
become clear upon computing the operator $\overline\partial$ and analyzing the space 
of solutions of eq. (6.8). 

Consider a local superfiled $X_\Xi$ of the form \xxx
$$
X_\Xi=\sum_{p,q\geq 0}\hbox{$1\over p! q!$}
{\Xi}^{a_1\ldots a_p}{}_{b_1\ldots b_q}(x)
y_{a_1}\cdots y_{a_p}dx^{b_1}\cdots dx^{b_q},
\eqno(6.9)
$$
where $\Xi\in\oplus_{p,q\geq 0}C^\infty(\wedge^p TM\otimes
\wedge^q T^* M\otimes\Bbb C)$ is a formal sum of biantisymmetric complex tensor 
fields on $M$ of varying bidegree $(p,q)$. Then, from (4.14$a$,$b$), 
through a tedious but totally straightforward computation, one finds 
the following expression: 
$$
\eqalignno{
\vphantom{1\over 2}
\overline\partial X_\Xi
&=\sum_{p,q\geq 0}\hbox{$1\over p!q!$}
\overline\partial_M \Xi^{a_1\ldots a_p}{}_{b_1\ldots b_q}(x)
y_{a_1}\cdots y_{a_p}dx^{b_1}\cdots dx^{b_q}&(6.10)\cr
\vphantom{1\over 2}
&+\sum_{p,q\geq 0}\hbox{$1\over p!q!$}
K_M{}^a \Xi^{a_1\ldots a_p}{}_{b_1\ldots b_q}(x)
dy_a y_{a_1}\cdots y_{a_p}dx^{b_1}\cdots dx^{b_q}, &\cr
}
$$
where
$$
\eqalignno{
\vphantom{1\over 2}
\overline\partial_M \Xi^{a_1\ldots a_p}{}_{b_1\ldots b_q}&=
\hbox{$1\over 2$}\Big\{(-1)^p q\Big[\partial_{[b_1}
\Xi^{a_1\ldots a_p}{}_{b_2\ldots b_q]}
+\sqrt{-1}\Big(J^c{}_{[b_1}\partial_{|c|}
\Xi^{a_1\ldots a_p}{}_{b_2\ldots b_q]}&(6.11a)\cr
\vphantom{1\over 2}
&-p(\partial_c J^{[a_1}{}_{[b_1}-\partial_{[b_1} J^{[a_1}{}_{|c|})
\Xi^{|c|a_2\ldots a_p]}{}_{b_2\ldots b_q]}
&\cr
\vphantom{1\over 2}
&-(q-1)\partial_{[b_1} J^c{}_{b_2}\Xi^{a_1\ldots a_p}{}_{|c|b_3\ldots b_q]}
\Big)\Big]
%&\cr
%\vphantom{1\over 2}
%&
-p\sqrt{-1}\Big[P^{[a_1|c|}\partial_c \Xi^{a_2\ldots a_p]}{}_{b_1\ldots b_q}
&\cr
\vphantom{1\over 2}
&-\hbox{$1\over 2 $}(p-1)\partial_c P^{[a_1a_2}
{\Xi}^{|c|a_3\ldots a_p]}{}_{b_1\ldots b_q}
+q\partial_{[b_1}P^{[a_1|c|}\Xi^{a_2\ldots a_p]}{}_{|c|b_2\ldots b_q]}\Big]
&\cr
\vphantom{1\over 2}
&+\hbox{$1\over 2 $}q(q-1)\sqrt{-1}\Big[
\partial_c Q_{[b_1b_2}+\partial_{[b_1} Q_{b_2|c|}+\partial_{[b_2}Q_{|c|b_1}
\Big]
\Xi^{ca_1\ldots a_p}{}_{b_3\ldots b_q]}\Big\},
&\cr
\vphantom{1\over 2}
K_M{}^a \Xi^{a_1\ldots a_p}{}_{b_1\ldots b_q}&=
\hbox{$1\over 2$}\Big\{\big(\delta^a{}_c+\sqrt{-1}J^a{}_c\big)
{\Xi}^{ca_1\ldots a_p}{}_{b_1\ldots b_q}
+(-1)^p\sqrt{-1}P^{ac}
{\Xi}^{a_1\ldots a_p}{}_{cb_1\ldots b_q}\Big\},
&\cr
&
&(6.11b)\cr
}
$$
the brackets $[\cdots]$ denoting full antisymmetrization of all enclosed 
indices except for those between bars $|\cdots|$.

At first glance, it would appear that eq. (6.8) is equivalent to the equations 
$$
\eqalignno{
\vphantom{1\over 2}
\overline\partial_M \Xi^{a_1\ldots a_p}{}_{b_1\ldots b_q}&=0,&(6.12a)\cr
\vphantom{1\over 2}
K_M{}^a \Xi^{a_1\ldots a_p}{}_{b_1\ldots b_q}&=0,&(6.12b)\cr
}
$$
at least for $q\leq 2$. This is indeed the case, though there are some 
subtleties involved. The $K_M{}^a \Xi^{a_1\ldots a_p}{}_{b_1\ldots b_q}$ are
manifestly the components of a tensor field. Eq. (6.12$b$) is therefore fully 
covariant. Conversely, as is straightforward to check, the 
$\overline\partial_M {\Xi}^{a_1\ldots a_p}{}_{b_1\ldots b_q}$ are not,
as they do not transform covariantly under coordinate changes. 
\footnote{}{}\footnote{${}^2$}{Note, however, that the combination of the two terms in 
the right hand side of (6.10) is fully covariant.} So, eq. (6.12$a$) is not covariant
in itself. However, it can be checked that, when restricting on the subspace 
of tensors $\Xi$ satisfying the covariant constraint (6.12$b$), 
$\overline\partial_M {\Xi}^{a_1\ldots a_p}{}_{b_1\ldots b_q}$ do transform 
covariantly and, thus, they are the components of a tensor field.
For such $\Xi$, eq. (6.12$a$) is therefore fully covariant. 

Suppose first our generalized complex structure is an ordinary 
complex structure $J$, so that $P=0$, $Q=0$ (cf. eq. (2.17)). 
By (6.11$b$), eq. (6.12$b$) implies that $\Xi$ is a formal sum of
$q$--forms with values in the holomorphic vector bundles $\wedge^pT^{1,0}M$
with varying $(p,q)$. Further, by inspection of 
(6.11$a$), one recognize $\overline\partial_M$ as the customary nilpotent 
Dolbeault operator on the space of these tensor fields $\Xi$. 
Hence, in this special case, for fixed $p$, eqs. (6.12$a$,$b$) 
define the customary Dolbeault cohomology with values in $\wedge^pT^{1,0}M$,
$H^{*,*}(M,\wedge^pT^{1,0}M)$. For this reason, we claim that, in the general 
case, eqs. (6.12$a$,$b$) define a notion of generalized Dolbeault cohomology.
The claim will be substantiated next. 

Denote by ${\scri X}^*$ the subspace of $\oplus_{p,q\geq 0}C^\infty(\wedge^p TM\otimes
\wedge^q T^* M\otimes\Bbb C)$ spanned by those $\Xi$ satisfying the constraint 
(6.12$b$). ${\scri X}^*$ is a graded vector space: for $n\in \Bbb Z$, 
${\scri X}^n$ is the subspace of ${\scri X}^*$ contained 
in $\oplus_{p,q, p+q=n}C^\infty(\wedge^p TM\otimes\wedge^q T^* M\otimes\Bbb C)$.

Exploiting (2.15$a$--$d$), by a very lengthy algebraic verification, one 
finds that
$$
\eqalignno{
\vphantom{1\over 2}
K_M{}^a\overline\partial_M&=0 \qquad \hbox{on ${\scri X}^*$},&(6.13a)\cr
\vphantom{1\over 2}
\overline\partial_M{}^2&=0 \qquad \hbox{on ${\scri X}^*$}.&(6.13b)\cr
}
$$
By relation (6.13$a$), ${\scri X}^*$ is invariant under 
$\overline\partial_M$. Relation (6.13$b$), in turn, states that 
$\overline\partial_M$ squares to $0$ on ${\scri X}^*$. 
Further, $\overline\partial_M$ maps ${\scri X}^n$ into ${\scri X}^{n+1}$, 
as is easy to check, and, so, has degree $1$.
Thus, the pair $({\scri X}^*,\overline\partial_M)$ is a cochain 
complex, with which there is associated a cohomology 
$H^*({\scri X}^*,\overline\partial_M)$. For reasons explained in the previous
paragraph, we call this generalized Dolbeault cohomology of $M$.

${\scri X}^*$ is actually a graded algebra and $\overline\partial_M$
is a derivation on this algebra. As a consequence, 
$H^*({\scri X}^*,\overline\partial_M)$ has an obvious ring structure.

It is easy to see that {\it eq. (6.9) defines a homomorphism of the 
generalized Dolbeault cohomology into the $\overline\partial$ superfield 
cohomology}. Recall that the latter is embedded in the classical 
Batalin--Vilkoviski cohomology. Therefore, 
{\it the classical Batalin--Vilkoviski cohomology is related non trivially to
\it the generalized Dolbeault cohomology of the target manifold $M$.}

The above analysis generalizes verbatim to the twisted case,
by replacing the Batalin--Vilkovisky operator $\delta_{BV}$ by its twisted 
counterpart $\delta_{BVH}$ with $H$ a closed $3$--form (cf. eqs (5.10$a$, $b$)). 
Only the explicit expression 
of the twisted generalized Dolbeault operator $\overline\partial_{MH}$ is different,
$$
\eqalignno{
\vphantom{1\over 2}
\overline\partial_{MH}\Xi^{a_1\ldots a_p}{}_{b_1\ldots b_q}&=
\overline\partial_{M}\Xi^{a_1\ldots a_p}{}_{b_1\ldots b_q}
-\hbox{$1\over 2$}\sqrt{-1}q\Big\{(-1)^p p H_{cd[b_1}P^{d[a_1}
\Xi^{|c|a_2\ldots a_p]}{}_{b_2\ldots b_q]}~~~~~~~~&(6.14)\cr
\vphantom{1\over 2}
&+\hbox{$1\over 2 $}(q-1)\Big[
H_{cd[b_1}J^d{}_{b_2}-H_{cd[b_2}J^d{}_{b_1}\Big]
\Xi^{ca_1\ldots a_p}{}_{b_3\ldots b_q]}\Big\},
}
$$
where the first term in the right hand side is given by (6.11$a$).
Using(2.23$a$--$d$), one can verify that relations (6.13$a$,$b$) still hold
with $\overline\partial_M$ replaced by $\overline\partial_{MH}$. 
So, a twisted cochain complex $({\scri X}^*,\overline\partial_{MH})$ 
and an associated twisted generalized Dolbeault cohomology 
$H^*({\scri X}^*,\overline\partial_{MH})$ can be defined. 
Expectedly, {\it this is closely related to the twisted classical Batalin--Vilkovisky 
cohomology.}

M. Gualtieri suggested to us that the generalized Dolbeault cohomology found above 
could be related to the cohomology of the deformation complex of untwisted 
generalized complex structure \ref{5}. We have found out that this is indeed the case, 
as we now show. 

%For simplicity, we consider first the untwisted case. 
Let ${\scri D}^*$ be the subspace of $\oplus_{p,q\geq 0}C^\infty(\wedge^p TM\otimes
\wedge^q T^* M\otimes\Bbb C)$ spanned by those $\Xi$ satisfying the constraint 
(6.12$b$) and the further constraint
$$
L_{Ma}\Xi^{a_1\ldots a_p}{}_{b_1\ldots b_q}=0,
\eqno(6.15)
$$
where \xxx
$$
L_{Ma}\Xi^{a_1\ldots a_p}{}_{b_1\ldots b_q}=
\hbox{$1\over 2$}\Big\{(-1)^p\big(\delta^c{}_a-\sqrt{-1}J^c{}_a\big)
{\Xi}^{a_1\ldots a_p}{}_{cb_1\ldots b_q}
+\sqrt{-1}Q_{ac}
{\Xi}^{ca_1\ldots a_p}{}_{b_1\ldots b_q}\Big\}.
\eqno(6.16)
$$
Then, ${\scri D}^*\subseteq {\scri X}^*$.
An algebraic verification similar to that yielding to (6.13$a$) shows that 
$L_{Ma}\overline\partial_M=0$ on ${\scri D}^*$. Thus, 
$({\scri D}^*,\overline\partial_M)$ is a subcomplex of 
$({\scri X}^*,\overline\partial_M)$. 

Next, suppose that we perform a {\it complex} deformation of the generalized complex 
structure $\cal J$ of the form 
$$
J'^a{}_b=J^a{}_b+\Xi^a{}_b, \qquad P'^{ab}=P^{ab}+\Xi^{ab}, \qquad
Q'_{ab}=Q_{ab}+\Xi_{ab},
\eqno(6.17)
$$
where $\Xi^a{}_b$, $\Xi^{ab}$, $\Xi_{ab}$ are the components of some element
$\Xi\in \oplus_{p,q, p+q=2}C^\infty(\wedge^p TM\otimes\wedge^q T^*M\otimes\Bbb C)$. 
Now, (6.17) defines a complex generalized almost complex 
structure ${\cal J}'$ to linear order in $\Xi$ provided $\Xi\in{\scri D}^2$. 
Indeed, imposing (2.13$a$--$c$) to first order in $\Xi$ yields the equations \xxx
$$
\eqalignno{
\vphantom{1\over 2}K_M{}^a\Xi_b+L_{Mb}\Xi^a&=0, &(6.18a)\cr
\vphantom{1\over 2}K_M{}^a\Xi^b+K_M{}^b\Xi^a&=0, &(6.18a)\cr
\vphantom{1\over 2}L_{Ma}\Xi_b+L_{Mb}\Xi_a&=0, &(6.18a)\cr
}
$$
which hold if $\Xi\in{\scri D}^2$. It is straightforward though lengthy to show 
that ${\cal J}'$ is integrable to linear order in $\Xi$ if $\Xi$ is a 2--cocycle of 
$({\scri D}^*,\overline\partial_M)$. For, imposing (2.15$a$--$d$) to first 
order in $\Xi$ and using (6.11$a$), (6.12$b$), (6.15)
yields a set of equations for $\Xi$ that can be cast as 
$$
\eqalignno{
\vphantom{1\over 2} 
&\overline\partial_M\Xi^{abc}=0,&(6.19a)\cr
\vphantom{1\over 2} 
&\overline\partial_M\Xi^{ab}{}_c=0,&(6.19b)\cr
\vphantom{1\over 2} 
&\overline\partial_M\Xi^a{}_{bc}=0,&(6.19c)\cr
\vphantom{1\over 2}
&\overline\partial_M\Xi_{abc}=0,&(6.19d)\cr
}
$$
so that $\overline\partial_M\Xi=0$. Finally, suppose that $\Upsilon\in {\scri D}^1$ 
and that $\Xi=\overline\partial_M\Upsilon$, so that 
$$
\eqalignno{
\vphantom{1\over 2} 
\Xi^{ab}&=\overline\partial_M\Upsilon^{ab},&(6.20a)\cr
\vphantom{1\over 2} 
\Xi^a{}_b&=\overline\partial_M\Upsilon^a{}_b,&(6.20b)\cr
\vphantom{1\over 2} 
\Xi_{ab}&=\overline\partial_M\Upsilon_{ab}.&(6.20c)\cr
}
$$
By explicit computation using (6.11$a$), (6.12$b$), (6.15), one finds
$$
\eqalignno{
\vphantom{1\over 2} 
\Xi^{ab}&=\hbox{$1\over 2$}\sqrt{-1}l_XP^{ab},&(6.21a)\cr
\vphantom{1\over 2} 
\Xi^a{}_b&=\hbox{$1\over 2$}\sqrt{-1}\big\{l_XJ^a{}_b
-P^{ac}(\partial_c\xi_b-\partial_b\xi_c)\big\},&(6.21b)\cr
\vphantom{1\over 2} 
\Xi_{ab}&=\hbox{$1\over 2$}\sqrt{-1}\big\{l_XQ_{ab}
+J^c{}_a(\partial_c\xi_b-\partial_b\xi_c)
-J^c{}_b(\partial_c\xi_a-\partial_a\xi_c)\big\},&(6.21c)\cr
}
$$
where $X^a=\Upsilon^a$  $\xi_a=\Upsilon_a$. Thus, $\Xi$ is a combination of a 
complex infinitesimal diffeomorphism $X$ and a complex
infinitesimal $b$ transform with $b=d_M\xi$
(cf. eqs. (2.14$a$--$c$)) and, so, represents a trivial deformation \ref{5}. 
This conclusively shows that $({\scri D}^*,\overline\partial_M)$ 
can be identified with the deformation complex of 
generalized complex structures. 

The above analysis can be carried out also in the twised case without much further 
effort. Since the deformation theory for the twisted case was not carried out in 
\ref{5}, we cannot perform any comparison. 

Further clarification of these matters should be left to the mathematicians. 
For us, it is sufficient to have found a remarkable connections between 
the Batalin--Vilkovisky cohomology of the Hitchin sigma model and 
various aspects of Hitchin's generalized complex geometry. 
             
\vskip.6cm\par\noindent{\bf Acknowledgments.}  
We thank F. Bastianelli for support and encouragement 
and M. Gualtieri for correspondence.

\vfill\eject

\vskip.6cm
\centerline{\bf REFERENCES}
\vskip.3cm

\item{[1]}
C.~Vafa,
``Superstrings and topological strings at large N'',
J.\ Math.\ Phys.\  {\bf 42} (2001) 2798,
\item{}
arXiv:hep-th/0008142.
%%CITATION = HEP-TH 0008142;%%

\item{[2]}
S.~Kachru, M.~B.~Schulz, P.~K.~Tripathy and S.~P.~Trivedi,
``New supersymmetric string compactifications'',
JHEP {\bf 0303} (2003) 061,
\item{} arXiv:hep-th/0211182.
%%CITATION = HEP-TH 0211182;%%

\item{[3]}
S.~Gurrieri, J.~Louis, A.~Micu and D.~Waldram,
``Mirror symmetry in generalized Calabi-Yau compactifications'',
Nucl.\ Phys.\ B {\bf 654} (2003) 61, \item{}
arXiv:hep-th/0211102.
%%CITATION = HEP-TH 0211102;%%

\item{[4]}
N.~Hitchin, 
``Generalized Calabi Yau manifolds'',
Q. \ J. \ Math. {\bf 54} no. 3 (2003) 281, 
\item{}
arXiv:math.dg/0209099.

\item{[5]}
M.~Gualtieri,
``Generalized complex geometry'',
Oxford University doctoral thesis,
\item{}
arXiv:math.dg/0401221.

\item{[6]}
E.~Witten,
``Mirror manifolds and topological field theory'',
in ``Essays on mirror manifolds'', ed. S.~T. ~Yau, International
Press, Hong Kong, (1992) 120,
arXiv:hep-th/9112056.
%%CITATION = HEP-TH 9112056;%%

\item{[7]}
A.~Kapustin,
``Topological strings on noncommutative manifolds'',
IJGMMP {\bf 1} nos. 1 \& 2 (2004) 49,
\item{}
arXiv:hep-th/0310057.
%%CITATION = HEP-TH 0310057;%%

\item{[8]}
S.~Fidanza, R.~Minasian and A.~Tomasiello,
``Mirror symmetric SU(3)-structure manifolds with NS fluxes'',
\item{}
arXiv:hep-th/0311122.
%%CITATION = HEP-TH 0311122;%%

\item{[9]}
M.~Grana, R.~Minasian, M.~Petrini and A.~Tomasiello,
``Supersymmetric backgrounds from generalized Calabi-Yau manifolds'',
\item{}
arXiv:hep-th/0406137.
%%CITATION = HEP-TH 0406137;%%

\item{[10]}
O.~Ben-Bassat,
``Mirror symmetry and generalized complex manifolds'',
\item{}
arXiv:math.ag/0405303.
%%CITATION = MATH-AG 0405303;%%

\item{[11]}
M.~Zabzine,
``Geometry of D-branes for general N=(2,2) sigma models'', 
\item{}
arXiv:hep-th/0405240.
%%CITATION = HEP-TH 0405240;%%

\item{[12]}
C.~Jeschek,
``Generalized Calabi-Yau structures and mirror symmetry'',
\item{}
arXiv:hep-th/0406046.
%%CITATION = HEP-TH 0406046;%%

\item{[13]}
A.~Kapustin and Y.~Li,
``Topological sigma-models with H-flux and twisted generalized complex manifolds'',
arXiv:hep-th/0407249.
%%CITATION = HEP-TH 0407249;%%

\item{[14]}
S.~Chiantese, F.~Gmeiner and C.~Jeschek,
``Mirror symmetry for topological sigma models with generalized Kahler geometry'',
\item{}
arXiv:hep-th/0408169.
%%CITATION = HEP-TH 0408169;%%

\item{[15]}
U.~Lindstrom,
``Generalized N = (2,2) supersymmetric non-linear sigma models'',
Phys.\ Lett.\ B {\bf 587} (2004) 216, 
\item{}
arXiv:hep-th/0401100.
%%CITATION = HEP-TH 0401100;%%

\item{[16]}
U.~Lindstrom, R.~Minasian, A.~Tomasiello and M.~Zabzine,
``Generalized complex manifolds and supersymmetry'',
\item{}
arXiv:hep-th/0405085.
%%CITATION = HEP-TH 0405085;%%

\item{[17]}
N.~Ikeda,
``Two-dimensional gravity and nonlinear gauge theory'',
Annals Phys.\  {\bf 235} (1994) 435, 
\item{}
arXiv:hep-th/9312059.
%%CITATION = HEP-TH 9312059;%%

\item{[18]}
P.~Schaller and T.~Strobl,
``Poisson structure induced (topological) field theories'',
Mod.\ Phys.\ Lett.\ A {\bf 9} (1994) 3129, 
\item{}
arXiv:hep-th/9405110.
%%CITATION = HEP-TH 9405110;%%

\item{[19]}
A.~S.~Cattaneo and G.~Felder,
``A path integral approach to the Kontsevich quantization formula'',
Commun.\ Math.\ Phys.\  {\bf 212} (2000) 591, 
\item{}
arXiv:math.qa/9902090.
%%CITATION = MATH-QA 9902090;%%

\item{[20]}
L.~Baulieu, A.~S.~Losev and N.~A.~Nekrasov,
``Target space symmetries in topological theories I'',
JHEP {\bf 0202} (2002) 021,
\item{}
arXiv:hep-th/0106042.
%%CITATION = HEP-TH 0106042;%%

\item{[21]}
I.~A.~Batalin and G.~A.~Vilkovisky,
``Gauge Algebra And Quantization'',
Phys.\ Lett.\ B {\bf 102} (1981) 27.
%%CITATION = PHLTA,B102,27;%%

\item{[22]}
I.~A.~Batalin and G.~A.~Vilkovisky,
``Quantization Of Gauge Theories With Linearly Dependent Generators'',
Phys.\ Rev.\ D {\bf 28} (1983) 2567
(Erratum-ibid.\ D {\bf 30} (1984) 508).
%%CITATION = PHRVA,D28,2567;%%

\item{[23]}
J.~Gomis, J.~Paris and S.~Samuel,
``Antibracket, antifields and gauge theory quantization'',
Phys.\ Rept.\  {\bf 259} (1995) 1,
\item{} arXiv:hep-th/9412228.
%%CITATION = HEP-TH 9412228;%%

\item{[24]}
T. Courant,
``Dirac manifolds'',
Trans. \ Amer. \ Math. \ Soc. {\bf 319} no. 2 (1990) 631.
  
\item{[25]}
T. Courant and A. Weinstein,
``Beyond Poisson structures'',
in ``Action hamiltoniennes des groupes, troisi\`eme th\'eor\`eme de Lie'',
Lyon (1986) 39, Travaux en Cours 27, Hermann, Paris (1988). 

\item{[26]}
R.~Zucchini,
``Target space equivariant cohomological structure of the Poisson sigma model'',
J.\ Geom.\ Phys.\  {\bf 48} (2003) 219,
\item{}
arXiv:math-ph/0205006.
%%CITATION = MATH-PH 0205006;%%

\item{[27]}
I.~Vaisman,
``Lectures on the Geometry of Poisson Manifolds'',
Progress in Mathematics, vol. 118, Birkh\"auser Verlag, Basel 1994.

\item{[28]}
C.~Klimcik and T.~Strobl,
``WZW-Poisson manifolds'',
J.\ Geom.\ Phys.\  {\bf 43} (2002) 341,
\item{}
arXiv:math.sg/0104189.
%%CITATION = MATH-SG 0104189;%%

\bye

\item{[]}
S.~J.~.~Gates, C.~M.~Hull and M.~Rocek,
``Twisted Multiplets And New Supersymmetric Nonlinear Sigma Models'',
Nucl.\ Phys.\ B {\bf 248}, 157 (1984).
%%CITATION = NUPHA,B248,157;%%

\item{[]}
M.~Kontsevich,
``Deformation quantization of Poisson manifolds'',
\item{}
arXiv:q-alg/9709040.
%%CITATION = Q-ALG 9709040;%%

\item{[]}
A.~S.~Cattaneo and G.~Felder,
``Poisson sigma models and deformation quantization'',
Mod.\ Phys.\ Lett.\ A {\bf 16} (2001) 179,
\item{}
arXiv:hep-th/0102208.
%%CITATION = HEP-TH 0102208;%%

\item{[]}
A.~S.~Cattaneo and G.~Felder,
``On the AKSZ formulation of the Poisson sigma model'',
Lett.\ Math.\ Phys.\  {\bf 56} (2001) 163, 
\item{}
arXiv:math.qa/0102108.
%%CITATION = MATH-QA 0102108;%%

\item{[]}
P.~Schaller and T.~Strobl,
``Poisson sigma models: A generalization of 2-d gravity Yang-Mills  systems'',
\item{}
arXiv:hep-th/9411163.
%%CITATION = HEP-TH 9411163;%%

\item{[]}
P.~Schaller and T.~Strobl,
``A Brief Introduction to Poisson sigma-models'',
\item{}
arXiv:hep-th/9507020.
%%CITATION = HEP-TH 9507020;%%

%%%%%%%%%%%%%%%%%%%%%%%%%%%%%%%%%%%%%%%%%%%

$$
\eqalignno{
\vphantom{1\over 2} 
\overline\partial_M\Upsilon^{ab}&=\hbox{$1\over 2$}\sqrt{-1}\Big\{
\Upsilon^c\partial_c  P^{ab}-\partial_c\Upsilon^a P^{cb}
-\partial_c\Upsilon^b P^{ac}\Big\},
&(6.20a)\cr
\vphantom{1\over 2} 
\overline\partial_M\Upsilon^a{}_b&=\hbox{$1\over 2$}\sqrt{-1}\Big\{
\Upsilon^c\partial_c J^a{}_b-\partial_c\Upsilon^a J^c{}_b
+\partial_b\Upsilon^c J^a{}_c-P^{ac}(\partial_c \Upsilon_b 
-\partial_b\Upsilon_c)\Big\},&(6.20b)\cr
\vphantom{1\over 2} 
\overline\partial_M\Upsilon_{ab}&=\hbox{$1\over 2$}\sqrt{-1}\Big\{
\Upsilon^c\partial_c Q_{ab}+\partial_a\Upsilon^c Q_{cb}
+\partial_b\Upsilon^c Q_{ac}+J^c{}_a(\partial_c \Upsilon_b 
-\partial_b\Upsilon_c)&(6.20c)\cr
&\hphantom{=\hbox{$1\over 2$}\sqrt{-1}\Big\{}-J^c{}_b(\partial_c \Upsilon_a 
-\partial_a\Upsilon_c)\Big\}.&\cr
}
$$

%%%%%%%%%%%%%%%%%%%%%%%%%%%%%%%%%%%%%%

$$
\eqalignno{
\vphantom{1\over 2}
\overline\partial_M {\Xi}^{a_1\ldots a_p}{}_{b_0\ldots b_q}&=
\hbox{$(-1)^p(q+1)\over 2 $}\Big[\partial_{[b_0}
{\Xi}^{a_1\ldots a_p}{}_{b_1\ldots b_q]}
+\sqrt{-1}J^c{}_{[b_0}\partial_{|c|}
{\Xi}^{a_1\ldots a_p}{}_{b_1\ldots b_q]}~~~~~~~~~&(6.11a)\cr
\vphantom{1\over 2}
&-p\sqrt{-1}(\partial_c J^{[a_1}{}_{[b_0}-\partial_{[b_0} J^{[a_1}{}_{|c|}
+H_{cd[b_0}P^{d[a_1}){\Xi}^{|c|a_2\ldots a_p]}{}_{b_1\ldots b_q]}
&\cr
\vphantom{1\over 2}
&-q\sqrt{-1}\partial_{[b_0} J^c{}_{b_1}{\Xi}^{a_1\ldots a_p}{}_{|c|b_2\ldots b_q]}
\Big],&\cr
K_M {\Xi}^{aa_1\ldots a_{p-1}}{}_{b_1\ldots b_q}&=
\hbox{$1\over 2$}\big(\delta^a{}_c+\sqrt{-1}J^a{}_c\big)
{\Xi}^{ca_1\ldots a_{p-1}}{}_{b_1\ldots b_q},
&(6.11b)\cr
\vphantom{1\over 2}
\varpi_M {\Xi}^{a_0\ldots a_p}{}_{b_1\ldots b_q}&=
(p+1)\Big[-P^{[a_0|c|}\partial_c {\Xi}^{a_1\ldots a_p]}{}_{b_1\ldots b_q}
&(6.11c)\cr
\vphantom{1\over 2}
&+\hbox{$1\over 2 $}p\partial_c P^{[a_0a_1}
{\Xi}^{|c|a_2\ldots a_p]}{}_{b_1\ldots b_q}
-q\partial_{[b_1}P^{[a_0|c|}{\Xi}^{a_1\ldots a_p]}{}_{|c|b_2\ldots b_q]}\Big],
&\cr
\vphantom{1\over 2}
\varsigma_M {\Xi}^{a_1\ldots a_{p-1}}{}_{b_0\ldots b_{q+1}}&=
\hbox{$(q+1)(q+2)\over 2 $}\Big[
\partial_c Q_{[b_0b_1}+\partial_{[b_0} Q_{b_1|c|}+\partial_{[b_1}Q_{|c|b_0}
&(6.11d)\cr
\vphantom{1\over 2}
&-H_{cd[b_0}J^d{}_{b_1}+H_{cd[b_1}J^d{}_{b_0}\Big]
{\Xi}^{ca_1\ldots a_{p-1}}{}_{b_2\ldots b_{q+1}]},
&\cr
\vphantom{1\over 2}
L_M {\Xi}^{aa_1\ldots a_p}{}_{b_1\ldots b_{q-1}}&=(-1)^pP^{ac}
{\Xi}^{a_1\ldots a_p}{}_{cb_1\ldots b_{q-1}},
&(6.11e)\cr
}
$$

\bye
$$
\eqalignno{
\vphantom{1\over 2}
\overline\partial_{MH}\Xi^{a_1\ldots a_p}{}_{b_1\ldots b_q}=\,
&\overline\partial_{M}\Xi^{a_1\ldots a_p}{}_{b_1\ldots b_q}&(6.14)\cr
\vphantom{1\over 2}
&+\hbox{$1\over 2$}\Big\{-(-1)^p qp\sqrt{-1}H_{cd[b_1}P^{d[a_1}
\Xi^{|c|a_2\ldots a_p]}{}_{b_2\ldots b_q]}&\cr
\vphantom{1\over 2}
&\hphantom{+\hbox{$1\over 2$}\Big\{}+\hbox{$1\over 2 $}q(q-1)\sqrt{-1}\Big[
-H_{cd[b_1}J^d{}_{b_2}+H_{cd[b_2}J^d{}_{b_1}\Big]
\Xi^{ca_1\ldots a_p}{}_{b_3\ldots b_q]}\Big\}.
}
$$

\vfill\eject

%%%%%%%%%%%%%%%%%%%%%%%%%%%%%%%%%%%%%%%%%%%%%%%%%%%%%%%%%%%%%%%%%%%5

Consider eqs. (6.12c), (6.12e) and neglect the others. 
By (eq. (6.11e), eq. (6.12e) implies that $\Xi$ is a $p$--vector 
with values in the bundle $\wedge^q N_P\otimes\Bbb C$, where $N_P$ is the annihilator 
bundle of the tangent bundle of the symplectic leaves associated to the Poisson 
structure $P$ \ref{26}. Further, by (6.11c), $\varpi_M$ is the customary nilpotent 
Poisson--Lichnerowicz operator \ref{26}. Hence, it is conceivable that eqs. (6.12) 
define some natural subspace of the Poisson--Lichnerowicz cohomology
$H^*_{PL}(M,\wedge^q N_P\otimes\Bbb C)$.

%%%%%%%%%%%%%%%%%%%%%%%%%%%%%%%%%%%%%%%%%%%%%%%%%%%%%%%%%%%%%%%%%%%%%%%

In this way, we realize in an essentially different way the program 
outlined in ref. \ref{} and further developed in ref. \ref{}.

A natural framework for answering this 
is Hitchin's method based on $\Cliff(6,6)$ spinors \ref{}. As we
review later in more detail, these are simply formal sums of forms on the
manifold.
Existence on a manifold of a pure nowhere $\Cliff(6,6)$ spinor is equivalent to 
the existence of a SU(3,3) structure. 
(If the spinor is also closed, Hitchin calls these
manifolds {\it generalized \cy s}.) 
For a SU(3) structure, there are {\sl two} pure spinors which are
orthogonal and of unit norm. From this point of view it is natural to
conjecture that mirror symmetry between two SU(3) structure manifolds 
exchanges these two pure spinors. We can be more explicit if we compare 
this $\Cliff(6,6)$ spinor definition of SU(3) structure with the more usual one, 
existence of a two--form $J$ and three--form 
$\Omega$ obeying $J\wedge\Omega=0$ and $i\Omega\wedge\bar\Omega=(2J)^3/3!$.

In these terms the two pure spinors are $e^{i J}$ and $\Omega$. We can
actually multiply first spinor by $e^B$ leaving it pure \cite{hitchin}.
So what we are claiming is 
$$
e^{B+iJ}\longleftrightarrow \Omega.
$$
In the Calabi Yau case, this exchange is implicit in many applications of
mirror symmetry. 

For example, the even periods 
and the D--brane charge can be written using $e^{B+iJ}$, and its exchange
with $\Omega$ was used in mapping \cite{lyz} stringy--corrected DUY equations
\cite{mmms} to the special lagrangian 
condition; $e^{B+i J}$ was also used in formulating  the concept of
$\Pi$--stability \cite{dfr}.

The issue of extending mirror symmetry to compactifications with fluxes 
has been studied recently in \cite{kstt,vafa,glmw,fmt}. 
A first question is of course
within which class of manifolds this symmetry should be defined. A natural
proposal comes from the formalism of G--structures, recently used in many
contexts of compactifications with fluxes.
As shown in \cite{glmw,fmt}, 
mirror symmetry can be defined on manifolds of SU(3)
structure, thus generalizing the usual Calabi--Yau case. One of the points
which makes this symmetry non--trivial is that, as expected, 
geometry and NS flux mix in the transformation. On the contrary, RR 
fluxes are mapped 
by mirror symmetry into RR fluxes and  their transformation is
well-understood. However, for many reasons it would be better to have a 
formalism
that would incorporate geometrical data and fluxes in 
a natural way.
This paper is a step in that direction. We will propose to use {\it pure
  spinors} as a formalism to describe SU(3)--structure compactifications.

\par
Recently, Hitchin formulated the notion of generalized complex geometry, which,
at the same time extends and unifies the customary notions of complex and 
symplectic geometry \ref{}. Hitchin's ideas were further developed by Gualtieri 
\ref{}.

Recent work on compactification of superstring theory with NS and RR fluxes turned 
on suggests that this geometry might be relevant in low energy 
superstring physics \ref{}. 
In particular, generalized complex geometry seems to be particularly suitable 
for the study of mirror symmetry \ref{}.

%%%%%%%%%%%%%%%%%%%%%%%%%%%%%%%   super Dirac delta %%%%%%%%%%%%%%%%%%%%%%%%

The super delta distribution $\delta$ is given by
$$
\delta(z,\zeta;z',\zeta')=
\hbox{$1\over 2$}\delta^{0,2}_{\alpha'\beta'}(z;z')
\zeta^{\alpha'}\zeta^{\beta'} 
+\delta^{1,1}_{\alpha\beta'}(z;z')\zeta^\alpha\zeta^{\beta'}
+\hbox{$1\over 2$}\delta^{2,0}_{\alpha\beta}(z;z')
\zeta^\alpha\zeta^\beta,
\eqno()
$$
$\delta^{p,1-p}(z;z')$ being the usual delta distributions for forms
on $\Sigma$. For a superfield $\psi$
$$
\int_{\Pi T\Sigma}\mu'\,\delta(z,\zeta;z',\zeta')\psi(z',\zeta')=\psi(z,\zeta).
\eqno()
$$

%%%%%%%%%%%%%%%%%%%%%%%%% superchains %%%%%%%%%%%%%%%%%%%%%%%%%%%%

In the case where $\Sigma$ has a non empty boundary
$\partial\Sigma$, the above relations hold provided the component fields 
of the superfield obey suitable boundary conditions \ref{}.

%%%%%%%%%%%%%%%%%%%%%   anti brackets   %%%%%%%%%%%%%%%%%%%%%%%%%%%%%%%%%%%%%%%%%%%%

$$
\eqalignno{\vphantom{1\over 2} 
&\big(x^a(z,\zeta),x^b(z',\zeta')\big)=0,&()\cr
\vphantom{1\over 2}
&\big(x^a(z,\zeta),y_b(z',\zeta')\big)=\delta^a{}_b\delta(z,\zeta;z',\zeta'),&()\cr
\vphantom{1\over 2}
&\big(y_a(z,\zeta),y_b(z',\zeta')\big)=0,&()\cr
}
$$

%%%%%%%%%%%%%%%%%%%%%%%%%%%%%%%%%%%%%%%%%%%%%%%%%%%%%%%%%%%%%

The 
local form of $x$ is 
$$
x^i(z,\zeta)=t^i(x(z,\zeta)).
\eqno()
$$

$$
\partial_ab_{bc}+\partial_bb_{ca}+\partial_cb_{ab}=0.
\eqno()
$$

$$
\partial_aH_{bcd}-\partial_bH_{acd}+\partial_cH_{abd}-\partial_dH_{abc}=0
\eqno()
$$

\item{[]}
S.~Cordes, G.~W.~Moore and S.~Ramgoolam,
``Lectures on 2-d Yang-Mills theory, equivariant cohomology and 
topological field theories'',
Nucl.\ Phys.\ Proc.\ Suppl.\  {\bf 41} (1995) 184
\item{} arXiv:hep-th/9411210.
%%CITATION = HEP-TH 9411210;%%

\item{[]}
D.~Birmingham, M.~Blau, M.~Rakowski and G.~Thompson,
``Topological field theory''.
Phys.\ Rept.\  {\bf 209} (1991) 129.
%%CITATION = PRPLC,209,129;%%

\item{[]} 
S.~Ouvry, R.~Stora and P.~van Baal, 
``On the Algebraic Characterization of Witten Topological Yang--Mills Theory'',
Phys.\ Lett.\ B {\bf 220} (1989) 159.

\item{[]} 
V.~Mathai and D.~Quillen, 
``Superconnections, Thom Classes and Equivariant Differential Forms'', 
Topology {\bf 25} (1986) 85.

\item{[]} M.~F.~Atiyah and L.~Jeffrey, 
``Topological Lagrangians and Cohomology'', 
J.\ Geom.\ Phys.\ {\bf 7} (1990) 119.

\item{[]} 
W. Greub, S. Halperin and R. Vanstone,
``Connections, Curvature and Cohomology'', vols. II and III, 
Academic Press, New York 1973.

\item{[]}
E.~Witten,
``Topological Quantum Field Theory'',
Commun.\ Math.\ Phys.\  {\bf 117} (1988) 353.

\item{[]}
J.~Grabowski, G.~Marmo and A.~Perelomov,
``Poisson Structures: toward a Classification'',
Mod.\ Phys.\ Lett.\ A {\bf 8} (1993) 1719

\item{[]}
E.~K.~Sklyanin,
``Some Algebraic Structures Connected With The Yang-Baxter Equation'',
Funct.\ Anal.\ Appl.\  {\bf 16} (1982) 263
[Funkt.\ Anal.\ Pril.\  {\bf 16N4} (1982) 27].
%%CITATION = FAAPB,16,263;%%

\item{[]}
A.~A.~Kirillov,
``Unitary Representations of Nilpotent Lie Groups'',
Russian Math.\ Surveys {\bf 17} (1962) 53.

\item{[]}
B.~Kostant,
``Orbits, Symplectic Structures and Representation Theory'',
Proc.\ U.S.--Japan Seminar of Diff.\ Geom.\ in Kyoto, (1966) 77.

\item{[]}
J.~--M.~Souriau,
``Structure des Syst\`emes Dynamiques'',
Dunod, Paris 1970.

\bye